\documentclass[preprint]{aastex}

\newcommand{\kms}{km~s$^{-1}$}

\begin{document}

\title{The Role of Drag in the Energetics of Strongly Forced Exoplanet Atmospheres}

\author{Emily Rauscher\footnote{NASA Sagan Fellow}
  \\ \textit{Lunar and Planetary Laboratory, University of Arizona,
  \\ 1629 East University Blvd., Tucson, AZ 85721-0092, USA}
  \\ and
  \\ Kristen Menou
  \\ \textit{Department of Astronomy, Columbia University,
  \\ 550 West 120th St., New York, NY 10027, USA}}

\begin{abstract}

In contrast to the Earth, where frictional heating is typically negligible, we show that drag mechanisms could act as an important heat source in the strongly-forced atmospheres of some exoplanets, with the potential to alter the circulation.  We modify the standard formalism of the atmospheric energy cycle to explicitly track the loss of kinetic energy and the associated frictional (re)heating, for application to exoplanets such as the asymmetrically heated ``hot Jupiters'' and gas giants on highly eccentric orbits.  We establish that an understanding of the dominant drag mechanisms and their dependence on local atmospheric conditions is critical for accurate modeling, not just in their ability to limit wind speeds, but also because they could possibly change the energetics of the circulation enough to alter the nature of the flow.  We discuss possible sources of drag and estimate the strength necessary to significantly influence the atmospheric energetics.  As we show, the frictional heating depends on the magnitude of kinetic energy dissipation as well as its spatial variation, so that the more localized a drag mechanism is, the weaker it can be and still affect the circulation.  We also use the derived formalism to estimate the rate of numerical loss of kinetic energy in a few previously published hot Jupiter models with and without magnetic drag and find it to be surprisingly large, at 5-10\% of the incident stellar irradiation.

\end{abstract}

\section{Introduction}

While the field of atmospheric dynamics is well established and mature, we are currently watching its new expansion into the exotic regimes introduced by extrasolar planets.  Much of the same, decades-old theory can be fruitfully applied to exoplanet atmospheres, but important differences exist between these planets and those found in our solar system.  This may result in the need to adapt theories to a new context and check that basic assumptions are not applied where they are not valid.

The strongly-irradiated atmospheres of close-in, extrasolar gas giants (``hot Jupiters") differ from solar system giants in a few important ways: 1) the stellar heating of the atmosphere dominates over internal heating, 2) these planets are expected to be tidally locked into synchronous orbits, meaning that the irradiation pattern is strongly asymmetric, and 3) the long rotational (= orbital) period means that the Coriolis force will have a weaker effect in these atmospheres (compared to Jupiter or Saturn) and atmospheric features should exist on a planetary scale (for a comprehensive review see Showman et al. 2010).  One important issue that is currently being investigated is the role of friction or drag in these atmospheres.

Atmospheres in the solar system are largely inviscid and estimates for hot Jupiters also imply very high Reynolds numbers \citep[e.g.][]{Li2010}; however, there can be other important sources of friction or drag in an atmosphere.  In particular, \citet{Goodman} questioned whether sources of drag were being correctly accounted for in hot Jupiter numerical models, concerned that a lack of drag could allow the winds to accelerate to unphysical speeds.  On the Earth it is well known that surface drag has an important impact on the momentum budget of the atmospheric flow.  While hot Jupiter atmospheres will not experience drag from a solid surface, other relevant mechanisms may be at work (e.g., shocks\footnote{We are loosely referring to shocks as drag, in the sense that they transform bulk kinetic energy into heat.}, instabilities, and magnetic drag, which we review in more detail in Section~\ref{sec:sources}).  It may be that models without explicit friction are still valid, if physical sources of friction are weak (compared to the radiative forcing) and/or numerical dissipation dominates, but it is necessary to have good estimates of the strengths of these mechanisms to determine if that is the case.

In this paper we will discuss the importance of friction in the context of the energetics of atmospheric circulation.  We emphasize that drag can have multiple effects on the flow, namely as: 1) a momentum sink, or 2) as a possible source of the heating that drives the flow.  (Throughout this paper we will be using the terms friction and drag synonymously.  Whether it is frictional dissipation from microscopic interactions, or large-scale drag from other physical processes, the result will still be the irreversible conversion of kinetic energy into localized heating.)  While the concerns raised in \citet{Goodman} were related to the first point, in this paper we will show that the second point may be of comparable importance in terms of correctly modeling exotic exoplanet atmospheres.

In Section~\ref{sec:global} we review standard equations for the global energy balance in an atmosphere (from Pearce 1978) and explicitly account for heating from frictional dissipation.  We then discuss the contexts in which it is acceptable to neglect friction from the energetics (even if not from the momentum equation).  In Section~\ref{sec:hJ} we discuss the difference between these contexts and the case of a strongly-forced atmosphere, with a discussion of the drag mechanisms that have been proposed for hot Jupiters (\ref{sec:sources}).  We then present equations for calculating the local values of the various energies and their conversion rates in Section~\ref{sec:local}, and in Section~\ref{sec:models} use these to estimate whether frictional heating should have been explicitly accounted for in previous numerical models.  We briefly discuss how these concepts will apply to the special case of an eccentric, close-in gas giant planet in Section~\ref{sec:eccentric}.  Finally, in Section~\ref{sec:summary} we summarize our main points and emphasize the context in which friction may matter to the energetics of atmospheric circulation.

\section{The Standard Theory of Friction in Atmospheric Energetics} \label{sec:global}

The total energy in an atmosphere is divided between several forms:  the specific kinetic energy ($\frac{1}{2}v^2$, where $v$ is the three-dimensional velocity), the internal energy\footnote{The internal energy of a parcel of air can change through compositional or phase changes (e.g. evaporation or condensation), but in this analysis we assume a constant state.} ($c_vT$, where $c_v$ is the specific heat and $T$ is the local temperature), and the gravitational potential energy ($gz$, where $g$ is the gravitational acceleration and $z$ is altitude).  If the atmosphere is in hydrostatic equilibrium\footnote{This is a common assumption---and usually justifiably so---for most atmospheres.  For a discussion of hydrostatic equilibrium in hot Jupiter atmospheres see \citet{Showman2008}.  Throughout this work we will assume that the atmospheres we consider are in hydrostatic equilibrium.} ($dP/dz=-\rho g$, where $P$ is pressure and $\rho$ is density), the ideal gas law ($P=\rho R T$) and the relation between the specific gas constant and specific heats ($R=c_p-c_v$) can be used to show that the integrated internal and gravitational potential energies for a column of air are equal to the integrated total specific enthalpy ($c_p T$).  In studying the energetics of an atmosphere, we want to track the sources and sinks of potential energy\footnote{Note that throughout this paper we will be using the term ``potential energy" to refer to \emph{thermodynamic} potential energy (usually in the form of enthalpy).  For an atmosphere in hydrostatic equilibrium the gravitational acceleration is balanced by the vertical pressure gradient; it is not this component of the potential energy that leads to winds, but rather the horizontal gradients of potential energy.} (through heating and cooling), the kinetic energy loss (through frictional dissipation), and the adiabatic conversion between potential and kinetic energies.  This conversion occurs when parcels of fluid flow down gradients created by differential heating in the atmosphere.

The concept of ``available" potential energy (APE) was first formulated by \citet{Lorenz1955} as the component of energy in an atmosphere that is generated by heating and available for direct conversion into kinetic energy.  His definition was in reference to some idealized, APE=0, state with no horizontal temperature/pressure gradients.  \citet{Pearce} created a new formalism based on sources, sinks, and conversions, in which he recognized that APE is only generated through variations in heating, whereas uniform heating only creates ``unavailable'' potential energy (UPE).  This concept of APE has also been formulated as available exergy or enthalpy \citep{Dutton,Marquet}.  We find the formalism by \citet{Pearce} most useful for our purposes here and in this section will follow his derivations.  Given that these are generally unfamiliar concepts in astrophysics, we consider it useful to review them here.

If $X$ is a local quantity (per unit mass) in the atmosphere, its global average is denoted by $\hat{X}$ and calculated as:
\begin{equation}
\hat{X}=\frac{1}{\mathcal{M}}\int \rho X dV =\frac{1}{\mathcal{M}}\int \frac{\rho X}{\rho g} dxdydP =\frac{1}{g\mathcal{M}}\int X dx dy dP
\end{equation}
where $\mathcal{M}$ is the total mass of the atmosphere, $x$ and $y$ are the horizontal coordinates, and the vertical coordinate has been converted from altitude to pressure by assuming hydrostatic equilibrium and that the gravitational acceleration ($g$) is constant.

Following \citet{Pearce}, we can then write global expressions for the time derivatives of the specific total potential ($e_{tot}$) and kinetic ($e_K$) energies of the atmosphere:
\begin{equation}
d \hat{e}_{tot} / d t = -\hat{c} + \hat{q}
\label{eq:e}
\end{equation}
\noindent and
\begin{equation}
d \hat{e}_K / d t = \hat{c} - \hat{\mathrm{d}}
\label{eq:k}
\end{equation}
\noindent where $\hat{q}$ is the globally averaged heating rate, $\hat{c}$ is the rate of conversion from potential to kinetic energy, and $\hat{\mathrm{d}}$ is the rate of dissipation of kinetic energy.  (In principle the conversion could go in either direction, but the loss of kinetic energy through dissipation requires a positive net conversion from potential to kinetic energy.)  To evaluate what fraction of the total potential energy will be available for conversion into kinetic energy, we begin with the thermodynamic equation:
\begin{equation}
ds/dt = (c_p/T)(dT/dt)-R\omega /P=q/T
\label{eq:dsdt}
\end{equation}
\noindent where $s$ is the specific entropy and $\omega=dP/dt$.  The global average of this equation is:
\begin{equation}
\label{eq:globals}
d \hat{s}/d t = \widehat{\left(\frac{q}{T}\right)} = \widehat{\hat{q}\left(\frac{1}{T}\right)}+\widehat{q^* \left(\frac{1}{T}\right)^*}
\end{equation}
\noindent where $^*$ denotes the deviation from the global average.  We can then define a reference temperature, $T_r$, by $1/T_r \equiv \widehat{(1/T)}$ and multiply Equation~\ref{eq:globals} by $T_r$ to obtain:
\begin{equation}
T_r d \hat{s}/d t = \hat{q} + \widehat{q^*\left(\frac{T_r}{T}\right)^*} = d \hat{u}/d t
\label{eq:u}
\end{equation}
\noindent The quantity $T_r d \hat{s}/d t$ is, by definition, the change in energy associated with a globally \emph{uniform} change in entropy and therefore defines the time derivative of the specific ``unavailable" potential energy (UPE, or $u$).  The potential energy that is available for conversion to kinetic energy (the APE, or $a$) will be the difference between this and the entire potential energy of the atmosphere (from Equation~\ref{eq:e}):
\begin{equation}
d \hat{a}/d t = d \hat{e}_{tot}/d t -d \hat{u}/d t = -\hat{c} - \widehat{q^*(T_r/T)^*}
\label{eq:a}
\end{equation}
\noindent  Through this derivation \citet{Pearce} was able to identify the term $-\widehat{q^*(T_r/T)^*}$ (the global average encompasses all variables) as the efficiency with which heating acts as a source for APE rather than UPE, and see that it depends on the spatial correlation between non-uniform heating and a non-uniform temperature structure.\footnote{Throughout this analysis we are implicitly assuming time-averaged values, including the definition of the reference temperature, $T_r$, as is appropriate for atmospheres with minimal global-scale time variability.  However, these equations are also valid for instantaneous values and could be expanded to analyze the energetics of a time-varying heating and temperature structure.}

We now return to the equations for the total potential and kinetic energies (\ref{eq:e},~\ref{eq:k}) to see that a steady state atmosphere ($d \hat{e}_{tot} /d t = d \hat{e}_K/d t = 0$) requires:
\begin{equation}
\hat{q} = \hat{c} = \hat{\mathrm{d}}
\label{eq:balance}
\end{equation}
\noindent  In order to explicitly track the frictional heating due to the dissipation of kinetic energy, we separate the heating rate into frictional and non-frictional\footnote{To be more explicit, we make a distinction between frictional heating---the increase in molecular kinetic energy from frictional dissipation of the kinetic energy of the flow---and non-frictional heating, which includes all other sources of heating, including radiation, conduction, and latent heat release.  Typically the radiative heating dominates over other non-frictional sources.} components, $q = q_f + q_{\mathrm{rad}}$.  We can then equate the global loss of kinetic energy with the global frictional heating ($\hat{\mathrm{d}}=\hat{q_f}$, an irreversible conversion) and see that Equation~\ref{eq:balance} implies:
\begin{equation}
\hat{q}=\hat{q}_f,\ \hat{q}_{\mathrm{rad}}=0.
\label{eq:aveq}
\end{equation}
\noindent The expression $\hat{q}_{\mathrm{rad}}=0$ is simply a statement of global radiative balance.\footnote{By assuming that the total energy of the atmosphere is conserved ($d \hat{e}_{tot} /d t = 0$) we are neglecting the evolutionary cooling of the planet.  The gas giants in our solar system are not in global radiative balance, with all but Uranus radiating more flux than they receive from the Sun.  However, here we are considering strongly forced atmospheres, where the stellar irradiation dominates over the internal heat flux and the amount by which the planet is out of global radiative equilibrium should be negligible.} We can also write the heating deviation as a combination of radiative and frictional components:
\begin{equation}
q^*=q^*_{\mathrm{rad}} + q^*_f.
\label{eq:devq}
\end{equation}
Taken all together, these equations from \citet{Pearce} identify the sources of potential energy in a steady-state atmosphere in global radiative equilibrium:
\begin{itemize}
\item The only source of UPE is the mean frictional heating.
\item The source of APE is differential heating, both radiative \emph{and frictional}.
\end{itemize}

With these equations we can now clearly formulate the context in which frictional heating can be neglected from the atmospheric energetics.  If the friction is mostly uniform it will all go into generating unavailable potential energy and will not directly affect the circulation.  If there is a good estimate of the amount of non-uniform frictional heating, it can be compared to the radiative heating to determine whether it can be a significant source of APE.  In the case of the Earth, for example, it has long been known that variations in the frictional heating are well below 10\% of the variations in radiative heating \citep{Pearce}.  Thus in the Earth-modeling context the surface friction is an important momentum sink, but can safely be neglected from the energetics without altering the global circulation.  We emphasize this point: when the friction is weak, and mostly uniform, it is safe to neglect it from the atmospheric energetics without altering the global circulation.

\section{Extension to Strongly Forced Exoplanet Atmospheres} \label{sec:hJ}

We now turn our attention to extrasolar planets, in particular hot Jupiters.  While they should not have a solid surface (the main source of friction in a terrestrial atmosphere), there are several drag mechanisms that have been suggested to exist in hot Jupiter atmospheres.  Most of these possible processes are significantly non-uniform and may be very strong.  However, the radiative forcing is also intense and a careful comparison of the relative strengths is warranted. 

Before moving on to discuss these atmospheres in more detail, we first point out that there is a complication in defining the atmosphere of the planet, as separate from the interior.  Unlike terrestrial planets, there is no clear boundary at a solid surface, but instead a smoother transition between the irradiation-driven (upper) atmosphere, a deeper radiative region that is not directly heated by the stellar insolation, and the interior convective zone at pressures of hundreds or thousands of bars \citep[e.g.,][]{Fortney2010}.  This makes it more difficult to define a closed system and calculate global rates of energy generation and exchange, since there will be some interaction with the deeper levels.  Although the radiative and advective timescales increase dramatically as we descend deeper into the planet, there may be some nonzero net transfer of momentum or heat; for example, the downward transport of kinetic energy and subsequent return as an extra internal heat flux has been proposed as one mechanism for inflating planetary radii \citep{SG02}.  These exchange rates may be important for describing the evolution of a planet throughout its lifetime, but here we are making the implicit assumption that we are averaging over some time period much less than the age of the planet (and that global-scale variability is minimal, consistent with many numerical models, including the ones analyzed in Section~\ref{sec:models}).  

There is currently no good prescription for how to model interaction between the upper atmosphere and deepest layers of hot Jupiters.  Numerical simulations typically assume that the modeled atmosphere is a closed box, with energy and momentum conserved.  For this analysis we will likewise assume that the upper atmosphere is decoupled from the interior--that any exchange terms are close to zero and can be neglected.  In the case that any exchange with the interior is mostly uniform, then the APE generation should not be much effected.  However, given the strongly day-night asymmetric nature of hot Jupiters, it is worth considering this issue more carefully.  Once there is a better understanding of how best to treat atmosphere-interior coupling, we will want to extend the formalism presented here to include exchange terms.  

\subsection{Discussion of possible drag mechanisms} \label{sec:sources}

Although many drag mechanisms could be at work in a planet's atmosphere, here we review three main physical processes that have been identified as possibly producing significant drag on hot Jupiters: vertical shear instabilities \citep{Goodman,Li2010},  shocks \citep{DobbsDixon2010,RM10}, and magnetic drag \citep{Perna2010a}.  

While numerical simulations are able to capture large-scale instabilities\footnote{An example of this is the horizontal shear instability that is triggered in the shallow hot Jupiter model of \citet{MR09} and which prevents flow speeds from reaching large values.}, the computational limits on resolution for global circulation models could eliminate any instabilities that are triggered on sub-grid scales.  Additionally, models that assume hydrostatic equilibrium have no formal vertical acceleration (vertical velocity exists, but results from the continuity equation) and as such, these models are unable to capture vertical shear instabilities, which must be parameterized.  As an attempt to determine what kind of sub-grid parameterization needs to be included in hot Jupiter global circulation models, \citet{Li2010} perform a linear analysis and non-linear simulations of limited regions of a hot Jupiter atmosphere (two-dimensional, with finite extents in longitude and altitude).  In their simulations they see weak transient shocks, but the main source of kinetic energy dissipation is through recurring Kelvin-Helmholtz instabilities.  Although there is not yet a clear prescription for how to do so, it will be important to determine how to connect processes in localized and global simulations to correctly treat the full range of atmospheric scales on hot Jupiters.

Transonic flow is present in most numerical simulations of hot Jupiter atmospheres.  While transient shocks may exist in the atmosphere, many models contain a shock (or hydraulic jump\footnote{A hydraulic jump is the equivalent of a shock in an incompressible hydrostatically-balanced flow; mass and momentum fluxes are continuous across the jump, but the energy flux is not \citep[cf.][]{Landau}.  In models that use the primitive equations of meteorology (instead of the full Navier-Stokes equations), vertical sound waves are filtered out and horizontal ones may or may not be, depending on the boundary conditions imposed.  See \citet{Kalnay} or \citet{Vallis}.}) as a steady feature \citep{DobbsDixon2008,Showman2008,S09,DobbsDixon2010,RM10}.  This is a localized region of strong dissipation and heating that has a markedly different spatial structure than the radiative heating pattern, often appearing on the night side of the planet, a fact that may strongly affect the APE generation.  The primitive equation models of \citet{Showman2008,S09} and \citet{RM10} do not include shock-capturing schemes and the heating associated with this feature is consistent with adiabatic heating of the downward flow that is induced by the strong convergence, rather than from any shock-heating.  The models in \citet{DobbsDixon2010} do include artificial viscosity and so should be able to correctly treat any generation of APE from shocks.  In principle the differences between these models could demonstrate the impact of standing shocks on the energetics of the circulation, but there are other important differences between these modeling approaches and it is difficult to disentangle the physics.

Finally, recent work implies that the atmospheres of hot Jupiters may be weakly thermally ionized and subject to interaction with the planetary magnetic field \citep{Batygin2010,Perna2010a}.  In particular, the mostly neutral flow may be dragged by the interaction between free charges and the magnetic field.  This mechanism is strongly affected by geometry---both as relates to the orientation of the magnetic field and due to conductivity differences between the hot dayside and cold nightside---and \citet{Perna2010a} estimate that the amount of drag could be strong enough to significantly alter the momentum of the flow.  The drag is associated with an induced magnetic field, itself sustained by electric currents which will dissipate ohmically in a non-local way.  The result of this complex interaction between the fluid and the magnetic field is that the kinetic energy dissipated will not be locally deposited as heating.  Nevertheless, the energy may be returned to the flow in a way that alters the energetics of the circulation.  A full MHD simulation may be required to fully understand how this process can affect the atmospheric flow.

There are also numerical sources of dissipation, although these are usually not well characterized and will vary from code to code, depending on the exact scheme used.  A known issue with global simulations is the build up of noise on small scales and a common technique is to use hyperdissipation (or hyperdiffusion) to remove power from the smallest scales \citep[see][for a study of how this affects the kinetic energy spectrum of the flow]{Thrastarson2011}.  Hyperdissipation is applied in order to fix an artificial result of the numerical simulation and in principle should not have an erroneous effect on the APE generation.  However, any scheme that alters the wind and/or temperature profiles, as this does, has the potential to change both the physical dissipation, as well as the efficiency with which that would be turned into reheating.

\subsection{Local theory of energetics} \label{sec:local}

In order to evaluate the importance of various possible drag mechanisms in the context of strongly forced atmospheres we need expressions for the local energetics.  In this section we will borrow from the derivation of available potential enthalpy in \citet{Marquet}, but extended to explicitly track dissipation and frictional heating.

For a steady ideal flow, without any net heating/cooling or dissipation of kinetic energy, we can use Bernoulli's theorem to define a set of locally conserved energies \citep[see, e.g.][]{Vallis}:
\begin{equation}
\frac{D}{Dt}\left[\frac{1}{2} v^2 + h + \Phi \right] = 0
\end{equation}
\noindent where $D/Dt$ is the material derivative, $h=c_vT + P/\rho = c_p T$ is the specific enthalpy, and $\Phi$ is the gravitational potential energy ($=gz$ for a constant gravitational field).\footnote{In Section~\ref{sec:global} we were able to include the gravitational energy as part of the total potential energy through integration over the atmosphere, but since we are now defining local terms we must keep track of these components separately.}  Our local equations for the energetics of the circulation must include these energies, the conversion terms between them, and local sources and sinks that can alter the energy budget locally, but should be globally conserved.  Recognizing that not all of the enthalpy will be available for conversion to kinetic energy, we employ a local definition of APE as the available potential enthalpy and define it as $a \equiv h-T_r s$, where we have subtracted off the unavailable component (UPE, $u \equiv T_r s$) that will globally integrate to agree with our previous definition of UPE (Equation~\ref{eq:u}).  Given the definition for specific entropy ($s=c_p\ln[\theta/\theta_r]$, where the potential temperature is $\theta = T [P_0/P]^\kappa$, and $\kappa=R/c_p$), we can expand out the local definition of APE:
\begin{eqnarray}
a     & = & h  - T_r s   \nonumber \\
       &=& c_p T- c_p T_r \ln \left(\frac{T}{T_r} \left[\frac{P_r}{P}\right]^\kappa \right)  \nonumber \\
       &=& c_p T - c_p T_r \ln(T/T_r) + R T_r \ln(P/P_r). \label{eq:adef}
\end{eqnarray}
\noindent Values with a subscript $r$ are for some reference state with uniform atmospheric temperature, defined as:
\begin{equation}
\frac{1}{T_r} \equiv \int_{atm} \frac{1}{T} \frac{dm}{\mathcal{M}} \label{eq:Tr}
\end{equation}
\noindent with an implicit time-average over an appropriate interval (equivalent to the constant $T_r$ from Section~\ref{sec:global}).  The specific unavailable potential enthalpy is:
\begin{eqnarray}
u & = & T_r s \nonumber \\
	& = & c_p T_r \ln (T/T_r) - R T_r \ln (P/P_r) \label{eq:udef}.
\end{eqnarray}

For an atmosphere in hydrostatic equilibrium the specific kinetic energy is dominated by the horizontal winds and we write: $e_K=\frac{1}{2} v^2$, with the understanding that $v$ is only the horizontal velocity (along equipotentials).  Finally, we must also track the gravitational potential energy: $e_G=gz$.

We now derive expressions for the rates of creation, conversion, and dissipation of the various atmospheric energies.  With reference to its definition (Equation~\ref{eq:udef}), the material derivative of the unavailable specific enthalpy is simply:
\begin{equation}
\frac{Du}{Dt}=T_r \frac{Ds}{Dt} = \left(\frac{T_r}{T}\right) q, \label{eq:dudt}
\end{equation}
\noindent  while from the definition of specific available enthalpy (Equation~\ref{eq:adef}) and the thermodynamic equation (Equation~\ref{eq:dsdt}) we can write:
\begin{eqnarray}
\frac{D a}{Dt} &=& \frac{Dh}{Dt} -T_r \frac{Ds}{Dt} \nonumber \\
		     &=& c_p \frac{DT}{Dt} - \frac{T_r}{T} q \nonumber \\
                        &=& \omega \frac{RT}{P} + q - \frac{T_r}{T} q \nonumber \\
                        &=& \frac{1}{\rho} \left( \frac{\partial P}{\partial t} + \vec{v} \cdot \nabla P + w \frac{\partial P}{\partial z} \right) + \left(1-\frac{T_r}{T}\right) q \nonumber \\
                        &=& \frac{1}{\rho} \left( \frac{\partial P}{\partial t} + \vec{v} \cdot \nabla P \right) -wg + \left(1-\frac{T_r}{T}\right) q \label{eq:dadt}.                        
\end{eqnarray}
\noindent  Here we have expanded out $\omega \equiv DP/Dt = \partial P/\partial t + \vec{v} \cdot \nabla P + w (\partial P/\partial z)$ and used the hydrostatic assumption: $w (\partial P/\partial z) = - w \rho g$, where $w \equiv Dz/Dt$ is the vertical velocity. 

For the kinetic energy we take the dot product of the horizontal velocity and the standard horizontal momentum equation, $D\vec{v}/Dt = -(1/\rho) \vec{\nabla} P - \vec{f} \times \vec{v} + \vec{F}_{\mathrm{fric}}$, where $\vec{f}=(2 \Omega \sin \phi) \vec{k}$ is the Coriolis parameter ($\Omega$ is the planetary rotation rate and $\phi$ is latitude) in the vertical direction ($\vec{k}$), to get:
\begin{equation}
\frac{De_K}{Dt} = - \frac{1}{\rho} (\vec{v} \cdot \nabla P) - \mathcal{D}, \label{eq:ek1}
\end{equation}
\noindent  where we have defined $\mathcal{D}$ as the irreversible frictional dissipation of kinetic energy ($= - \vec{v} \cdot \vec{F}_{\mathrm{fric}}$).  The derivative of the gravitational potential energy is, by definition:
\begin{equation}
\frac{D(e_G)}{Dt} = g \frac{Dz}{Dt} = gw \label{eq:eg}
\end{equation}

We explicitly split the heating into its frictional ($q_f$) and non-frictional (mainly radiative, $q_{\mathrm{rad}}$) components and combine Equations~\ref{eq:dudt}-\ref{eq:eg} into a local energy cycle:\footnote{Note that the primary difference between this energy cycle and the one presented in \citet{Marquet} is that we have explicitly included the frictional heating as a source of potential energy.  This also necessitates the inclusion of UPE in our set of equations.}
\begin{eqnarray}
\frac{Du}{Dt} &=& \left[\frac{T_r}{T}\right] (q_{\mathrm{rad}} + q_f) \label{eq:finalu} \\
\frac{D a}{Dt} &=&   \left[1-\frac{T_r}{T}\right] (q_{\mathrm{rad}} + q_f) + \frac{1}{\rho} (\vec{v} \cdot \nabla P) -wg + \frac{1}{\rho}\frac{\partial P}{\partial t} \label{eq:finala}  \\
\frac{De_K}{Dt} &=&  - \frac{1}{\rho} (\vec{v} \cdot \nabla P) - \mathcal{D} \label{eq:finalke}  \\
\frac{De_G}{Dt} &=& wg \label{eq:finaleg} 
\end{eqnarray}
\noindent  The heating/cooling of the atmosphere ($q_{\mathrm{rad}}+q_f$) acts as a source/sink of APE and UPE, with the local temperature structure ($T_r/T$) acting as an efficiency factor for generating one form of potential energy or the other.  APE will be generated in regions of the atmosphere that are hotter-than-average ($T_r/T < 1$, so $[1-T_r/T] >0$) and heated ($q>0$) or cooler-than-average ($T_r/T > 1$) and cooled ($q<0$).  Note that the global integral of the APE source term ($[1-T_r/T]q$) is equal to the global source term derived previously in Equation~\ref{eq:a} ($-\widehat{q^* [\frac{T_r}{T}]^*}$).  The APE is converted to kinetic energy by the term $-(1/\rho)(\vec{v} \cdot \nabla P)$, which is equivalent to $\vec{v} \cdot \nabla (gz)$ if pressure is used as the vertical coordinate.  There will be exchange between APE and gravitational potential energy ($wg$), which relates to the puffing up of a hydrostatic atmosphere when it is heated.  The term $(1/\rho)(\partial P/\partial t)$ is related to adiabatic expansion and disappears under a steady state assumption.  

If dissipated kinetic energy is returned locally as frictional heating, so that $\mathcal{D}=q_f$, the sum of Equations (\ref{eq:finalu}-\ref{eq:finaleg}) will be $D(u+a+e_K+e_G)/Dt=q_{\mathrm{rad}}$, for a steady-state atmosphere.  While some of the processes described above may result in a local conversion of kinetic energy to heat, it is important to recognize that this will not always be the case.  One example is magnetic drag, which will induce a magnetic field, and the resulting current will be dissipated non-locally.  In this case the ohmic heating will not occur where the kinetic energy is dissipated and locally $\mathcal{D} \ne q_f$.  Non-local reheating can also result when instabilities trigger gravity waves, which then propagate through the atmosphere and deposit heat and momentum where they dissipate \citep{Watkins2010}.

We can now see that the importance of returning the dissipated kinetic energy as a source of localized heating will depend on its strength relative to the radiative heating ($q_f$ vs. $q_{\mathrm{rad}}$), but also on its spatial correlation with the efficiency factor $1-T_r/T$.  Given a drag mechanism with a known strength and spatial distribution (and assuming localized reheating), we can use these expressions to estimate its effect on the energetics of the circulation.

\subsection{Drag and energetics in previously published numerical models} \label{sec:models}

\citet{Perna2010a} present three numerical models that attempt to estimate the effect of magnetic drag on (the momentum of) hot Jupiter circulation.  Although magnetic interaction between the winds and any planetary magnetic field could result in significant drag, the effect will not be localized heating since the currents sustaining the induced magnetic field will be ohmically dissipated in a non-local way \citep[see][for studies of the related heating]{Batygin2010,Perna2010b}.  This means that we cannot use these models to estimate the amount of localized heating that results from magnetic drag.  However, in the spirit of simplicity, we use the same drag strength as the magnetic drag in these models, and we assume local dissipation, in order to explore the effect of frictional heating on the atmospheric energetics and calibrate the level at which it may become important.

We compare the models in \citet{RM10} and \citet{Perna2010a}, which all have identical set-ups for a generic hot Jupiter (see the papers for details), except for different amounts of Rayleigh drag applied as a term in the momentum equation.  The model in \citet{RM10} has no drag applied, while the three models in \citet{Perna2010a} have drag strengths that vary with height (stronger higher in the atmosphere).  Drag timescales range from $10^7-10^9$ seconds in the weakest-drag model (PMRa in Table 1), and are a factor of 10 and 100 shorter in the medium- and strongest-drag models (PMRb and PMRc in Table 1).  For reference, advective timescales are on the order of $10^5$ seconds and radiative timescales vary from $\sim3\times10^3$ to greater than $10^7$ seconds (increasing with depth).

The local specific kinetic energy loss is calculated from the prescribed drag timescales as $\mathcal{D} = (1/2) v^2/\tau_{\mathrm{drag}}$.  The radiative heating is treated by a Newtonian relaxation scheme, making it simple to determine the value of $q_{\mathrm{rad}}$ at each point in the modeled atmosphere.  We calculate the mass-weighted\footnote{The mass element for each point in the atmosphere is $dm=\rho dxdydz = - (1/g) dxdydP$.} global integrals of these quantities, as well as the source terms for UPE and APE (from Equations~\ref{eq:finalu} and~\ref{eq:finala}), and present our results in Table 1.  Note that our numerical solver does not convert the lost kinetic energy into reheating (in other words, while $\mathcal{D}$ is nonzero in Equation~\ref{eq:finalke}, $q_f=0$ in Equations~\ref{eq:finalu} and~\ref{eq:finala}).  The values listed for the generation of UPE and APE are $(T_r/T)q_{\mathrm{rad}}$ and $(1-T_r/T)q_{\mathrm{rad}}$, respectively, while the values listed as ``missing" generation replace $q_{\mathrm{rad}}$ with $q_f=\mathcal{D}$ to determine how much potential energy should have been generated if we did indeed include frictional heating in the models.

Before discussing the results, we address one issue of concern with these models, which is that the deepest levels (below $\sim$10 bar) may have not completely reached a statistically steady state by the end of the 1450 or 500 planet days simulated in \citet{RM10} and \citet{Perna2010a}.   It is likely that these deep levels are still being accelerated through interaction with the upper atmosphere.  This could complicate our energetics calculations and so we have rerun and extended all of the runs from those papers to mitigate this issue.  We calculate energy rates at 4975, 5000, and 10000 planet days for the \citet{RM10} model, and at 5000 days for the \citet{Perna2010a} models.  

\begin{deluxetable}{lcccc}
\tablewidth{0pt}
\tablecaption{Global energetics of various runs}
\tablehead{
\colhead{}  &  \colhead{RM10\tablenotemark{a}}   & \colhead{PMRa\tablenotemark{b}}  &  \colhead{PMRb\tablenotemark{b}} & \colhead{PMRc\tablenotemark{b}}
}
\startdata
Radiative heating, $q_{\mathrm{rad}}$ ($\times 10^{19}$ W)         & 160, 150, 180 & 390 & 430 & 610 \\
Generation of UPE ($\times 10^{19}$ W)                           & -39, -65, -20    & -19  & -45  & -5.0 \\
Generation of APE ($\times 10^{19}$ W)                           & 200, 210, 200 & 410 & 470 & 620 \\ \hline
Drag dissipation, $\mathcal{D}$ ($\times 10^{19}$ W)                           & 0                         & 3.8 & 34 & 170 \\
\ \ \ Relative to $q_{\mathrm{rad}}$:                                                      & 0                         & 1\% & 8\% & 28\% \\
Missing UPE gen. ($\times 10^{19}$ W)                         & 0                         &   7.3 & 68 & 340 \\
\ \ \ Relative to $q_{\mathrm{rad}}$ UPE gen.:                                   & 0                         & 38\% & 150\% & 6800\% \\
Missing APE gen. ($\times 10^{19}$ W)                         & 0                          &  -3.4 & -34 & -170 \\
\ \ \ Relative to $q_{\mathrm{rad}}$ APE gen.:                                   & 0                           & 1\% & 7\% & 27\% \\ \hline
Heat engine efficiency, $\eta$					& 7\%	& 13.7\%	& 15.7\%	& 20.7\% \\
\ \ \ adjusted for missing APE gen.:				& --		& 13.6\%	& 14.5\%	& 15.0\% \\
\enddata
\label{map_tab:one}
\tablenotetext{a}{From runs identical to the one presented in \citet{RM10}, except extended out to 4975, 5000, and 10000 planet days.}
\tablenotetext{b}{From runs in \citet{Perna2010a}, with increasing amounts of drag for PMRa, PMRb, and PMRc, associated with planetary magnetic field strengths of 3, 10, and 30 G, respectively.  These results are for simulations run for 5000 planet days.}
\tablecomments{All values are global totals, integrated over the atmosphere.  The dissipation is only what has been explicitly included in the model and does not account for any numerical dissipation.  In these models the dissipated kinetic energy is not re-included as localized heating, but the values we list as ``missing" generation show what the heating would have added to the global energies.}
\end{deluxetable}

\subsubsection{Global energy rates}

The first point worth noting is that for all of these models the global integral of $q_{\mathrm{rad}} \neq 0$, meaning that the planet is not in global radiative equilibrium.  There are two possible resolutions to this issue.  The first is that the atmosphere is still being accelerated by the heating and so has yet to reach a steady state.  However, if this were the case, the value of $q_{\mathrm{rad}}$ for the \citet{RM10} run at 10000 days should be less than at 5000 days, whereas there is not significant difference between them.  We have also inspected the time evolution of kinetic and internal energies throughout the runs and we see no evidence for continued acceleration of the flow or heating of the atmosphere.

The second explanation, which we prefer, is that the non-zero value of $q_{\mathrm{rad}}$ is a measure of the amount of numerical dissipation in our model runs.  As can be clearly seen in Equation~\ref{eq:balance}, any amount of dissipation (physical or numerical), $\hat{\mathrm{d}}$, requires a non-zero average global heating rate, $\hat{q}$, in order for there to be a net conversion from APE to kinetic energy to balance its loss.  While \emph{physically} the non-zero global heating rate will be supplied by the global frictional heating ($\hat{q}_f$ in Equation~\ref{eq:aveq}), in our numerical models the frictional heating is not included ($q_f=0$) and so the global radiative heating, $q_{\mathrm{rad}}$, must be non-zero to compensate for numerical losses ($\mathrm{d}_{\mathrm{num}}$).  This interpretation is supported by noting that $q_{\mathrm{rad}}$ is higher for the runs with drag applied, because in these cases the increased amount of dissipation (now both numerical and physical, $\mathrm{d}_{\mathrm{num}}+\mathrm{d}_{\mathrm{drag}}$) must be balanced by an increased amount of heating ($q_{\mathrm{rad}}$, given that $q_f$ remains zero).

In order to estimate the kinetic energy loss in our models, we can compare the global dissipation (assuming $\mathrm{d}=q_{\mathrm{rad}}$) to the input heating from incident stellar flux.  Our radiative forcing is applied through a Newtonian relaxation scheme \citep[for details see][]{RM10}, so that there is no well-defined stellar flux incident on the model atmosphere.  However, the values chosen for the forcing are derived from a one-dimensional radiative transfer model \citep{Iro05} for the hot Jupiter HD 209458b and so it is reasonable to compare our dissipation to a stellar heating rate appropriate for that planet.\footnote{The planet also has an internal heat flux, but that is $<$1\% of the stellar heating.}  Assuming an incident stellar heating rate of $\sim3 \times 10^{22}$ W, our drag-free model's $q_{\mathrm{rad}} \simeq 1.5 \times 10^{21}$ W ($=\mathrm{d}_{\mathrm{num}}$) implies a numerical dissipation rate of $\sim5$\%.  

For the models with an increasing amount of physical drag applied (PMRa, b, c) the global numerical dissipation is $\mathrm{d}_{\mathrm{num}}=q_{\mathrm{rad}}-\mathcal{D}$ and apparently dominates over the physical drag loss ($\mathcal{D}$), with a magnitude evaluated as $(q_{\mathrm{rad}}-\mathcal{D})/(3 \times 10^{22}) \simeq 13-15$\%.  Without a more detailed analysis it is difficult to know where this numerical loss of kinetic energy occurs in our model, be it evenly throughout the atmosphere or preferentially at regions of strong flow convergence or shear.  The amount of kinetic energy lost may also be dependent on the order and magnitude of hyperdissipation used.\footnote{All of the models shown here were run at T31 (spectral) resolution using hyperdissipation applied as an eighth-order operator on the vorticity, divergence, and temperature fields with a coefficient of $8.54\times10^{47}$ m$^8$ s$^{-1}$.}  These results imply that our numerical scheme leads to a significant amount of kinetic energy loss for the hot Jupiter atmospheric regime, an issue that will require further attention outside of this paper.

As expected, the amount of (physical) kinetic energy loss ($\mathcal{D}$) increases with the amount of drag applied to the atmosphere and it also becomes equal to a greater percentage of the radiative heating (although this is not linear with shorter drag timescales, since the increased drag will also work to decrease wind speeds and reduce the kinetic energy).  From the formalism presented above, we know that it is not the ratio of $q_f/q_{\mathrm{rad}}$ that matters in determining whether the energetics of the circulation will be altered, but the contribution that the frictional heating would make to the generation of APE (Equation~\ref{eq:finala}).

We find that in these models the spatial correlation between the atmospheric temperature structure and the kinetic energy dissipation is such that, in all cases, frictional heating would have worked to reduce APE generation, instead of feeding UPE.  In these models the effect of frictional heating from drag on the atmospheric flow should be twofold: it will work directly to reduce wind speeds \emph{and} the associated heating would also work to decrease the amount of APE generated and available to drive the winds.

The importance of this effect is determined by the ratio of the APE generation that would have come from frictional heating and the APE generation from radiative heating.  We find that for the amount of drag applied to models PMRa, PMRb, and PMRc, the relative importance of APE generation from frictional heating is 1, 7, and 27\%, respectively.  We have demonstrated that---for this spatial function for the drag---the amount of drag necessary to significantly affect the energetics of the circulation is something at a strength comparable to the medium-strength drag applied in the models of \citet{Perna2010a}.  We also know that the energetics would not have been significantly changed by the amount of drag applied to the weakest-drag model, while in the strongest-drag model the missing frictional heating would have significantly reduced the rate of APE generation.

The energetic effect of frictional heating can also be evaluated by considering the atmosphere as a heat engine and calculating its efficiency.  The standard Carnot efficiency is defined as $\eta = (T_W - T_C)/T_W$, where $T_W$ and $T_C$ are the temperatures of the warm source and cold sink, respectively.  We calculate $T_W$ and $T_C$ for each model as the average temperature of the regions of the atmosphere that are being diabatically heated and cooled, respectively, weighted by the local strength of the heating/cooling.  This gives $\eta=$ 3, 9, 12, and 17\% for RM10, PMRa, b, and c, respectively.  However, this is not the best estimate of the heat engine efficiency of the atmosphere.  For the case of the Earth, the Carnot efficiency gives a value of $\eta \sim 10\%$, but a better estimate comes from the actual energy input into the system: the ratio of the frictional dissipation ($=$ APE generation) to incoming solar flux, which gives $\eta \approx 0.8 \%$ \citep{Peixoto1992}.

If we compare the global APE generation to incoming stellar heating ($\approx 3 \times 10^{22}$ W) for our models we get $\eta \approx$ 7, 13.7, 15.7, and 20.7\% for RM10, PMRa, b, and c, respectively.  This indicates that, seen as heat engines, these hot Jupiter atmospheres are at least an order of magnitude more efficient than the Earth's atmosphere.  The efficiency increases with the amount of drag applied to the hot Jupiter atmosphere, as the same amount of radiative forcing has to be used more efficiently to generate APE, which feeds the kinetic energy, which is then dissipated at a higher rate.  Recalculating the efficiency that would have resulted from including the ``missing" APE generation, we get $\eta \approx$ 13.6, 14.5, and 15.0\% for the PMRa, b, and c models.  Since the frictional heating in these models is always working against APE generation, the efficiency is reduced.  The statement that frictional heating would have reduced the APE generation by X\% is equivalent to saying that the efficiency of the atmosphere as a heat engine would be reduced by X\%.

While this does provide an estimate of the strength of drag that may energetically affect the atmospheric flow---assuming local reheating from dissipation---we note that our result is dependent on the set-up we used, in particular the spatial form used for the drag.  As we have emphasized throughout, it is the spatial structure of the atmosphere that dictates the flow energetics, specifically the correlations between temperatures, heating rates, and dissipation rates.  Our results here are dependent on the particular relation between the spatial function we have used for the atmospheric drag and the underlying atmospheric structure.  Many of the drag mechanisms mentioned earlier will have very different spatial dependencies.  The best way to estimate the energetic importance of any drag mechanism is to apply the Equations~(\ref{eq:finalu}-\ref{eq:finaleg}) to a particular atmosphere with some form of drag at work.

\subsubsection{Spatial structure of the energetics}

We can further understand the energetics of the atmosphere by studying its spatial properties.  This will allow us to identify the regions of the atmosphere that are the most relevant for APE generation, as determined by the spatial correlation between the atmospheric quantities of interest.  

Before we proceed, we need to highlight an important nuance in this analysis.  Although Equations~(\ref{eq:finalu}) and (\ref{eq:finala}) give local expressions for UPE and APE generation, note that the local values only have meaning in reference to the global total.  The quantity $T_r$ is explicitly a global value and is an inherent part of the definitions of APE and UPE, both of which were originally motivated by the concept of a reference state with APE$=0$.  The reader should keep in mind that the following plots of APE and UPE generation are to be taken as an indication of which areas of the atmosphere (when integrated over) contribute the most to the global total.

\begin{figure}[ht!]
\begin{center}
\includegraphics[width=0.45\textwidth]{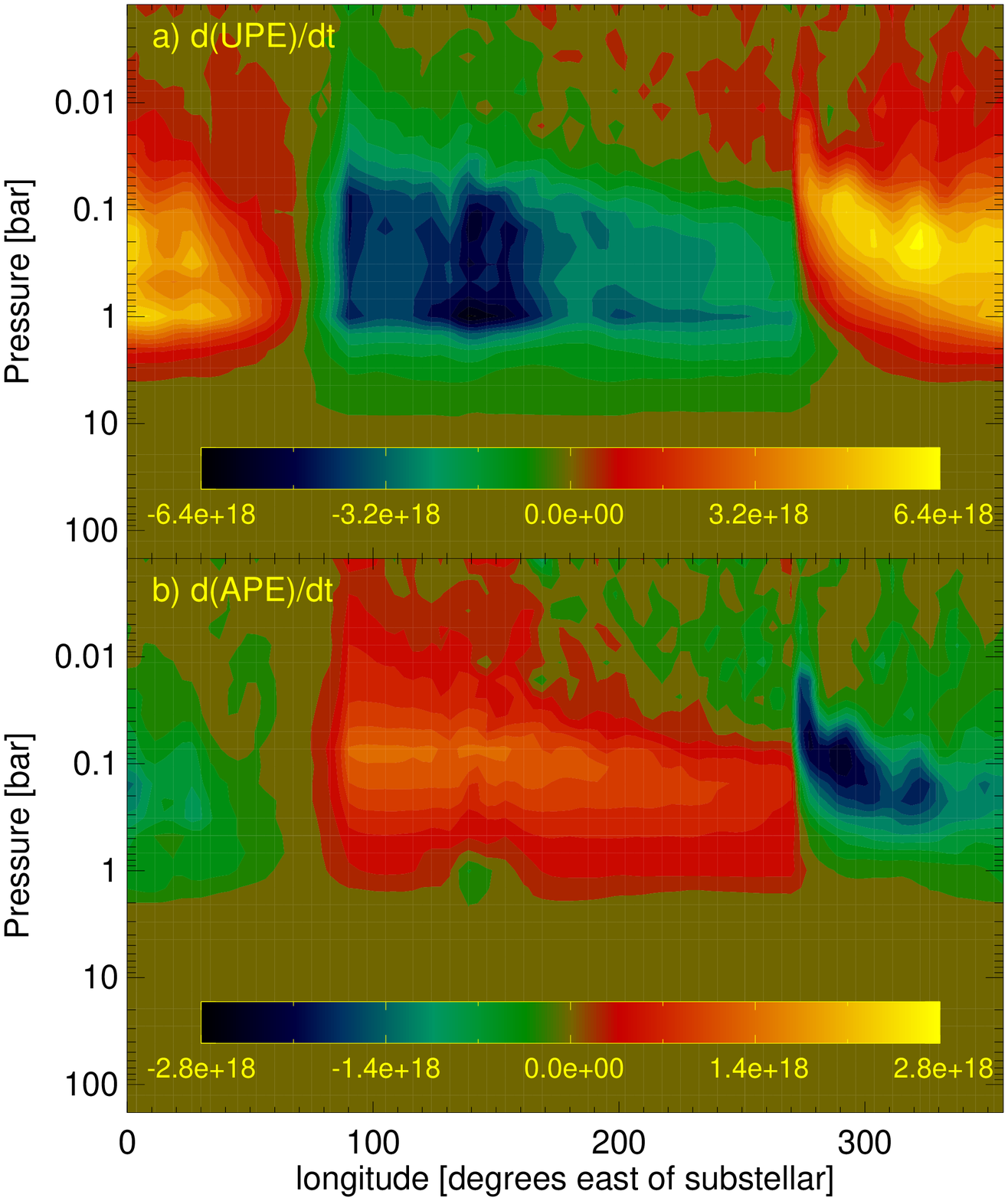}
\includegraphics[width=0.45\textwidth]{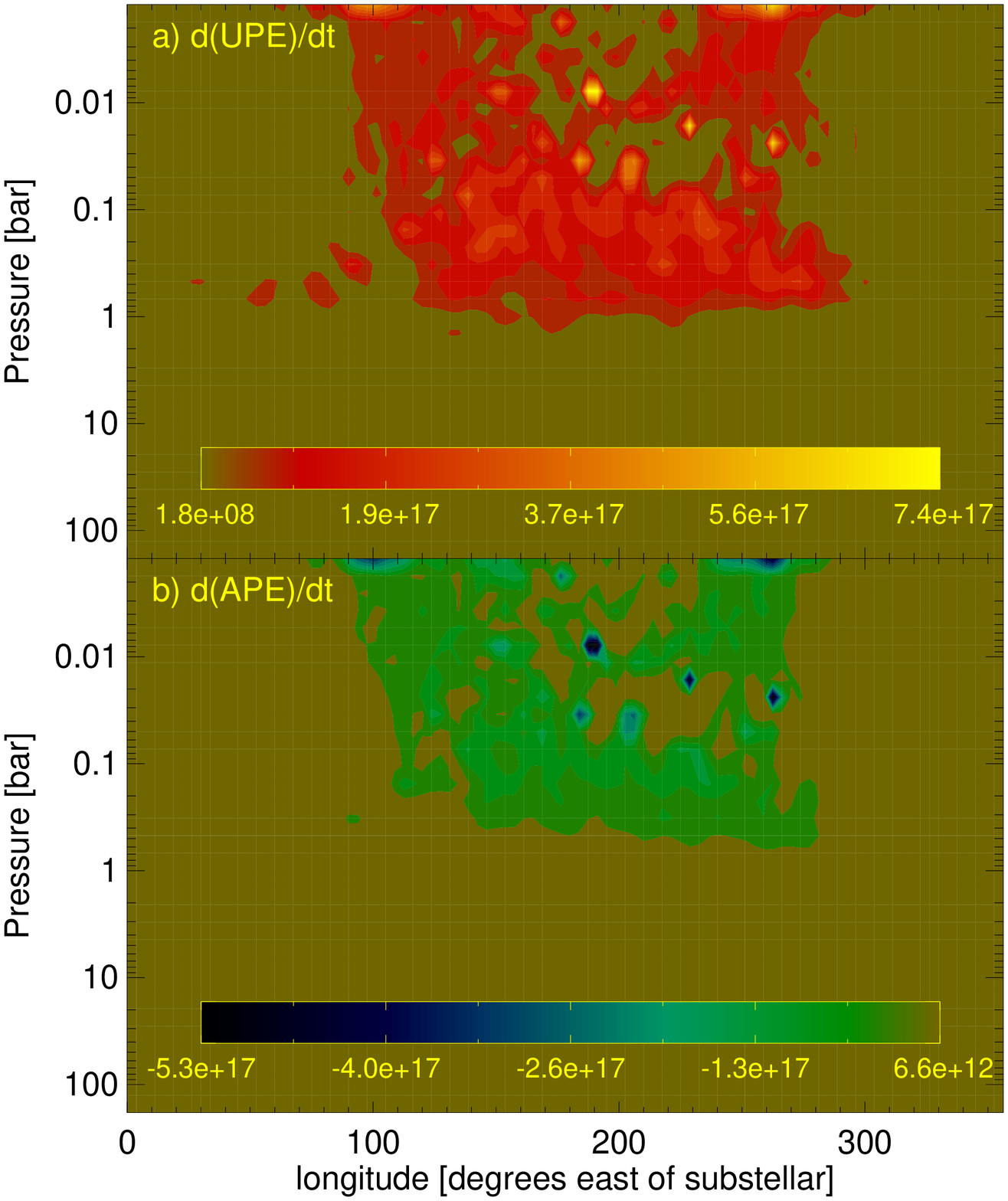}
\end{center}
\caption{Plots of the generation of available and unavailable potential energy (APE, UPE, in watts), for an equatorial slice through the atmosphere (the equator dominates the energy budget).  \emph{Left:} the UPE and APE generated/lost through radiative heating/cooling in the RM10 model.  \emph{Right:} for the strongest-drag model (PMRc), the UPE and APE that \emph{would} have been generated/lost through frictional heating, if it had been included and locally equal to the kinetic energy lost from drag.  The colors have been scaled so that red/yellow is always positive and green/blue is always negative.  See the text for further explanation.} \label{fig:eq}
\end{figure}

In Figure~\ref{fig:eq} we plot the local values of UPE and APE generation as a function of longitude and pressure (the model's vertical coordinate) for an equatorial slice through the atmosphere.  (The energy rates tend to be greatest at the equator.)  In the left panel we show the radiative generation for the \citet{RM10} model at 10,000 days, while in the right panel we plot the ``missing" frictional generation for the strongest-drag model from \citet{Perna2010a} at 5000 days.  The pattern of radiative UPE and APE generation is similar for all of the models, although the amplitudes and some of the detailed structure is different.  The ``missing" UPE and APE generation in the PMRa and PMRb models is weaker and primarily concentrated in the uppermost layers of the atmosphere, on the night side.  The longer drag times in these models mean that it is only the highest levels (where the applied drag is the strongest) which are able to significantly dissipate kinetic energy, even though most of the kinetic energy resides in the deeper layers.  (See Figure 4 of Rauscher \& Menou 2010 for a plot of kinetic energy versus depth; it has a maximum at $\sim2$ bar.)

\newpage

From Figure~\ref{fig:eq} we can see that \emph{locally} the radiative generation of UPE is much greater than the generation of APE, although we know from Table 1 that it will integrate to a smaller global rate.  This is a reflection of the fact that the radiative heating on hot Jupiters is intense and a large amount of energy will be stored in UPE, but by its very nature this will have a small net effect on the atmospheric flow.

According to the left panel of Figure~\ref{fig:eq}, most of the UPE is generated on the day side and most of the APE is generated on the night side (with opposite behavior for the loss).  This is simple to explain by comparing the signs of the local efficiency factor ($T_r/T$) and the radiative heating/cooling ($q_{\mathrm{rad}}$).  The calculation of the reference temperature, $T_r$, is a mass-weighted integral over the atmosphere (Equation~\ref{eq:Tr}) and the deepest layers, which contain the most mass, are generally hotter than the upper layers (see Figure 5 of Rauscher \& Menou 2010 for a plot of temperature-pressure profiles for different locations throughout the model).  For the models under consideration here, $T_r \simeq 1750-1800$ K,\footnote{$T_r$ is cooler for RM10 and hotter for the PMR runs.  There was a variation of $\sim$1\% in $T_r$ between the days sampled for RM10.} and $T_r/T >1$ for most of the atmosphere at pressures less than $\sim10$ bar.  This means that UPE/APE will be generated where the atmosphere is heated/cooled so that $q_{\mathrm{rad}}$ is greater/less than zero (see Equations~\ref{eq:finalu} and~\ref{eq:finala}).  In general cool air advected from the night side will be heated on the day side, while flow of hot air from day to night will result in cooling on the night side, although we will discuss deviations from this later.  The maximum UPE/APE generation will occur where the heating/cooling is strongest, weighted by $T_r/T$.  The amount of heating and cooling is a function of how much the winds are able to bring gas out of local radiative equilibrium and how quickly the gas can radiatively respond.  These properties are themselves both related to the generation of APE and conversion to kinetic energy.

The UPE and APE generation that would have resulted from localized frictional heating has a different spatial pattern than the radiative generation.  In the right panel of Figure~\ref{fig:eq} we see that the frictional UPE generation (and APE loss) is primarily on the night side.  The majority of the kinetic energy in the atmosphere is found on the night side and since we have adopted a horizontally uniform drag in these models, this results in having most of the dissipation on the night side.  If the drag were instead stronger on the day side, for example, we might see a more even pattern of dissipation around the planet.  Frictional heating will always be positive (the dissipation of kinetic energy does not lead to cooling) and so in the upper atmosphere (at $P < 10$ bar, where $T_r/T >1$) it will always lead to a decrease in APE.  In this set-up the night side has most of the radiative APE generation and most of the frictional APE loss.

As a test of how much our answers are affected by the value of the reference temperature, we recalculated $T_r$ by only integrating over the upper layers ($P<10$ bar), where the atmosphere is being radiatively heated.  This gave $T_r \approx 1600-1650$ K.  We then used this value to calculate the global energy rates (over the entire model atmosphere) and found that our radiative APE and UPE generation remained unchanged, while our ``missing" frictional generation was decreased by $\sim10-20$\%.  However, it is not appropriate to restrict the definition of $T_r$ to a partial region of the atmosphere.  Although not directly heated, the inert layers (at $P>10$ bar) contain a significant fraction of the total kinetic energy, which has been gained through interaction with the upper layers, and so must be included as part of the closed system.  The real issue is then where to set to the bottom boundary of the model, as discussed above at the beginning of Section~\ref{sec:hJ}.

\begin{figure}[ht!]
\begin{center}
\includegraphics[width=0.49\textwidth]{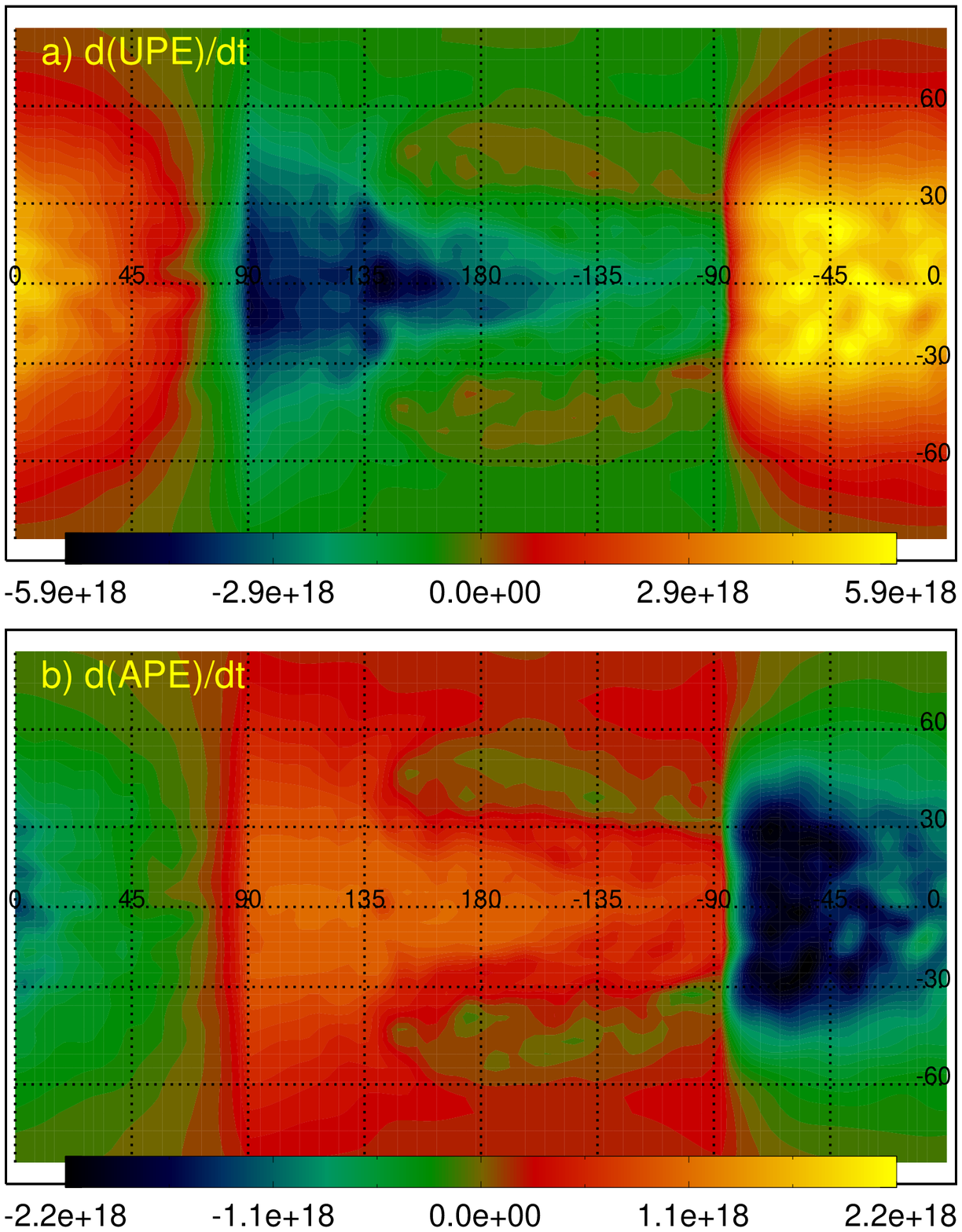}
\includegraphics[width=0.49\textwidth]{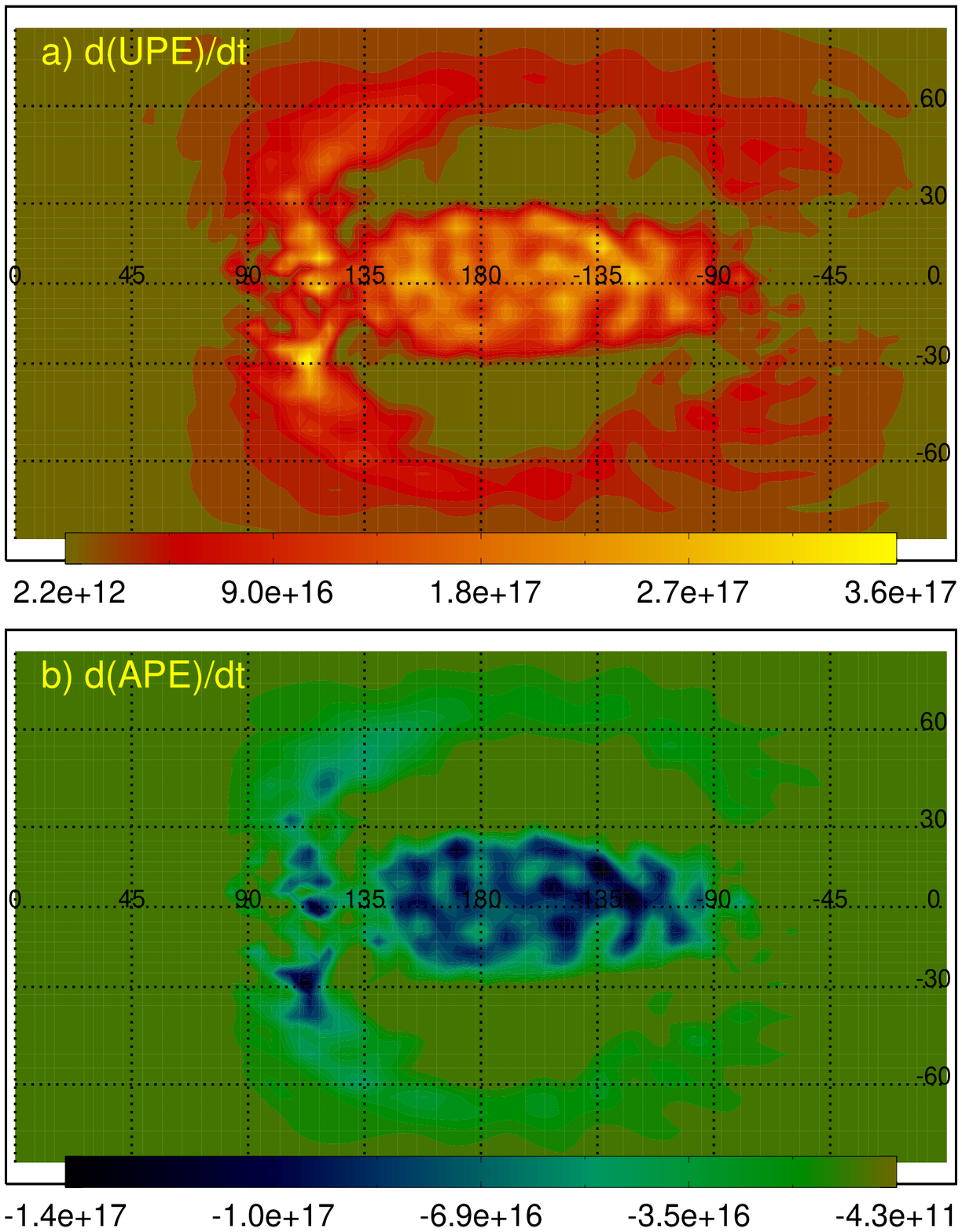}
\end{center}
\caption{Cylindrical projections of the generation of available and unavailable potential energy (APE, UPE, in watts), for a horizontal slice of the atmosphere at 150 mbar, with the substellar point at (0,0).  \emph{Left:} the UPE and APE generated/lost through radiative heating/cooling in the RM10 model.  \emph{Right:} for the strongest-drag model (PMRc), the UPE and APE that \emph{would} have been generated/lost through frictional heating, if it had been included and locally equal to the kinetic energy lost.  The colors have been scaled so that the red/yellow is always positive and the green/blue is always negative.  See text for further explanation.} \label{fig:150}
\end{figure}

We can also examine the horizontal structure of APE generation.  In Figure~\ref{fig:150} we plot the UPE and APE generation as a function of latitude and longitude at the 150 mbar pressure level.  The left panel is the radiative generation from the \citet{RM10} model at 10,000 days and the right panel is the ``missing" frictional generation for the strongest-drag model from \citet{Perna2010a} at 5,000 days.  The radiative generation at this level in the PMRc model is similar to what we see for the RM10 model, although with the amplitudes slightly decreased and the spatial structure more smoothed out.  We choose to show this particular level because it contains significant generation of potential energy, from both the radiative forcing and the ``missing" frictional heating.

\begin{figure}[ht!]
\begin{center}
\includegraphics[width=0.49\textwidth]{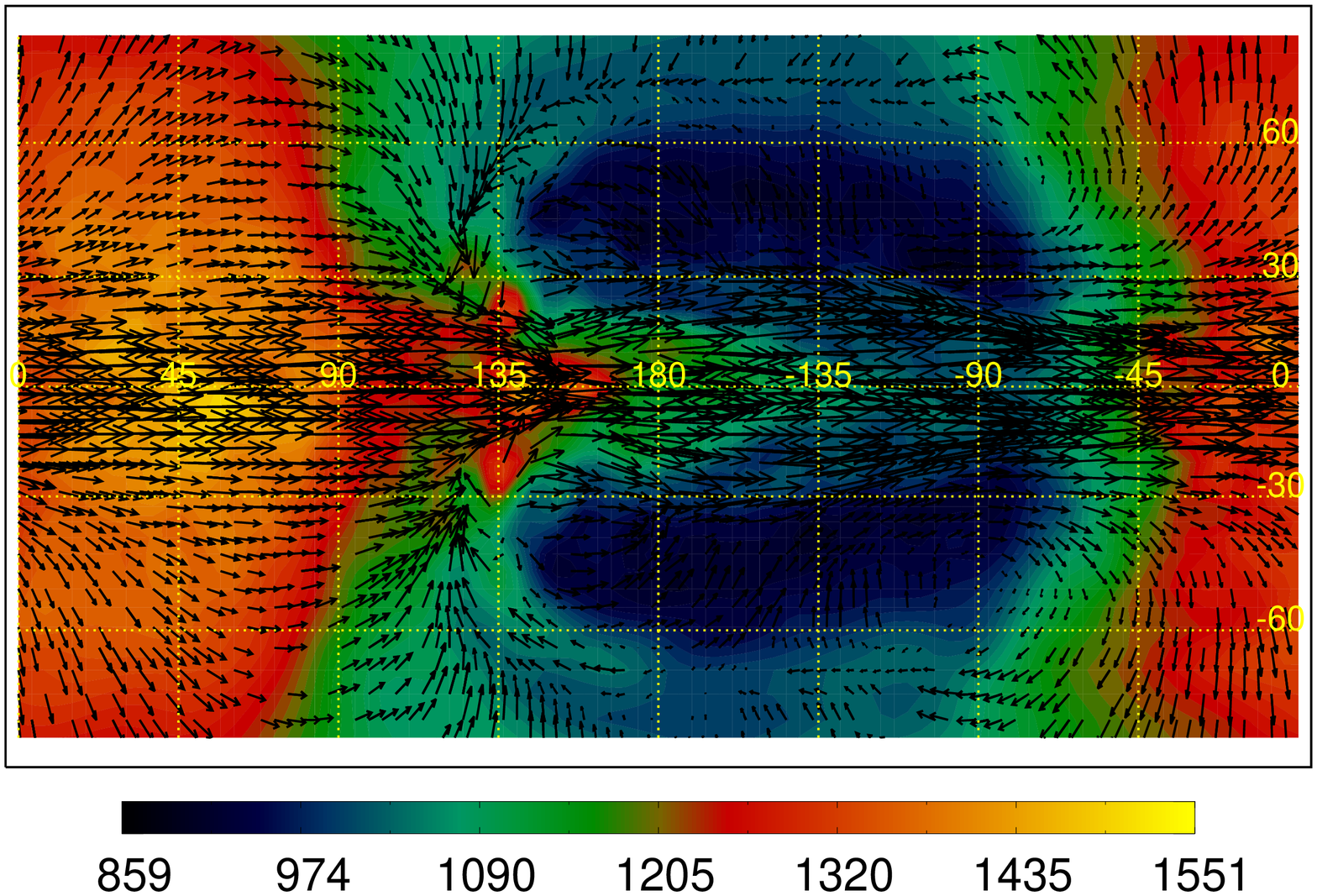}
\includegraphics[width=0.49\textwidth]{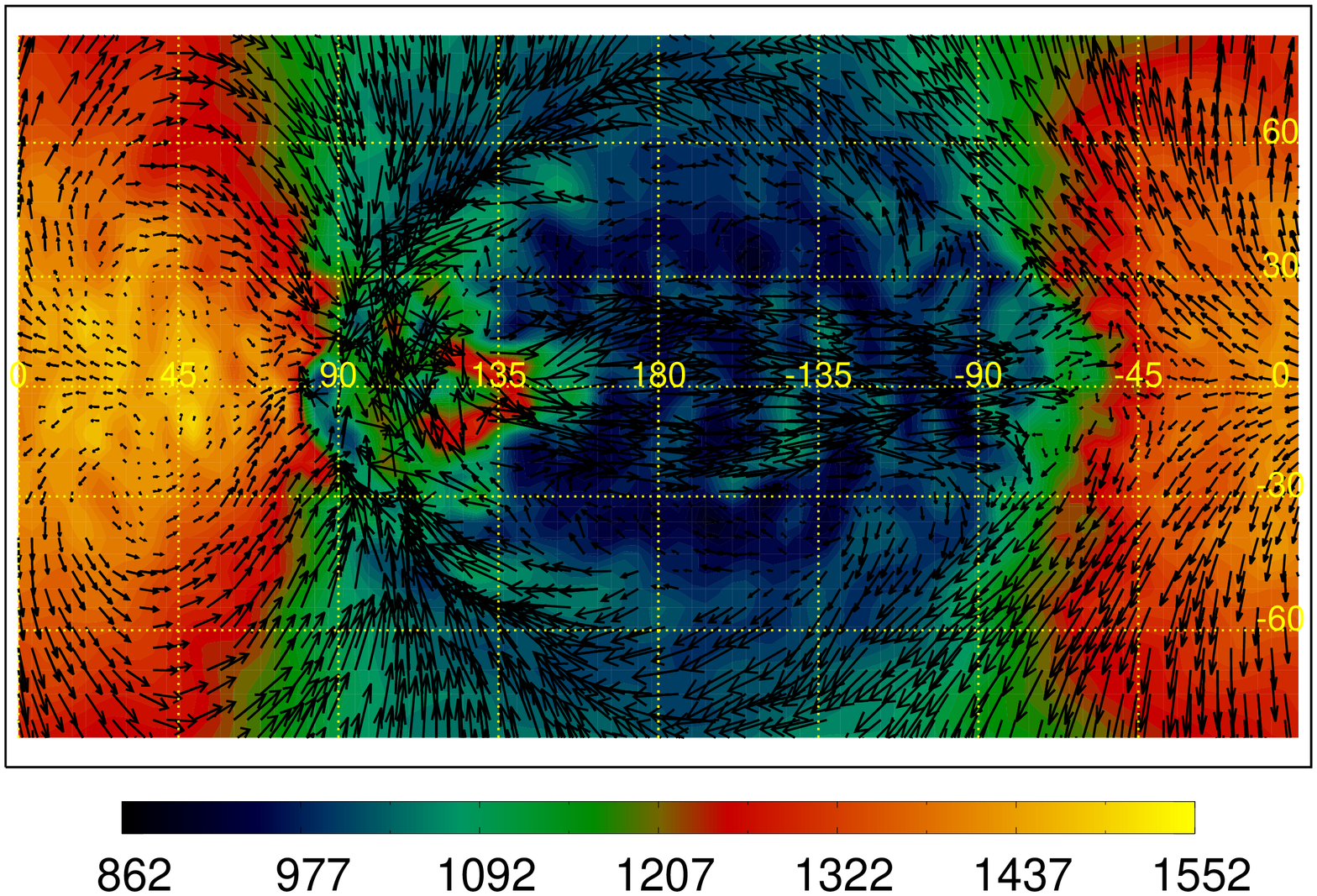}
\end{center}
\caption{Cylindrical projections of temperature (color, in K) and velocity (arrows, where length scales with magnitude) for a horizontal slice through the atmosphere at 150 mbar, with the substellar point at (0,0).  \emph{Left:} for the RM10 model.  Zonal (east-west) wind speeds range from -1.6 to 4.3 \kms~and meridional wind speeds range from -1.7 to 2 \kms.  \emph{Right:} for the PMRc model.  Zonal wind speeds range from -2.5 to 2.6 \kms~and meridional wind speeds range from -2.8 to 2.8 \kms.  In both models the sound speed at this level ranges from 2.1 to 2.9 \kms.} \label{fig:Tv}
\end{figure}

As we discussed above, the generation of potential energy depends on the temperature structure (by $T_r/T$), spatially correlated with the radiative or frictional heating rate.  The radiative heating generally works to make the day side hotter and the night side colder, with a dependence on the advected temperature structure, while the frictional heating depends on the flow structure and wind speeds.  In order to more clearly understand the patterns of potential energy generation, we plot the temperature and velocity structures for both models at this pressure level in Figure~\ref{fig:Tv}.

In agreement with Figure~\ref{fig:eq}, we see that most of the radiative UPE generation (and APE loss) is on the day side, while this is reversed on the night side.  This is explained, as before, by the fact that the efficiency factor, $T_r/T$, is greater than 1 at this level and so day side heating increases UPE while night side cooling increases APE.  The deviations from this pattern are due to the detailed structure of the advected temperature pattern, scaled by the local efficiency factor (at this level $T_r/T$ ranges between $\sim$2 in the coldest regions and $\sim$1.2 in the hottest regions).  In general we see eastward advection, so that hot air is pulled from day to night across the east terminator (at $90^\circ$) and cold air is pulled across the west terminator.  The boundary between positive and negative generation is not exactly at $90^\circ$ longitude because the air that was heated at the substellar point begins to cool even before it reaches the night side.  This effect is also seen at the west terminator, close to the poles, where the flow is westward.

The region of reversed sign on the night side (positive UPE, negative APE generation) at $\sim \pm45^\circ$ latitude is because the air there is cooler than the night side equilibrium temperature and so is being heated, as prescribed by our Newtonian relaxation scheme.  Note that night side heating is not unphysical.  Although there is no incident stellar light, there is still flux from the deeper, hotter atmospheric layers, which will prevent the night side temperatures from dropping too low.

We also find that the strongest rates of potential energy generation are not at the sub- and anti-stellar points ([$0^\circ$, $0^\circ$] and [$180^\circ$, $0^\circ$] in longitude and latitude, respectively).  On the day side the heating is primarily of cold air advected across the west terminator (-$90^\circ$) and at this level the relation between advective and radiative timescales is such that the maximum heating occurs at $\sim -45^\circ$.  On the night side there are two regions of cooling: where hot air has been advected from the day side across the east terminator, and where air has been heated by the shock-like feature at $\sim135^\circ$.  This feature is discussed in more detail in \citet{RM10}; it is a steady feature that extends across many pressure levels and is associated with hot regions, consistent with adiabatic heating from downward flow induced by the strong horizontal convergence.  Note that in Figure~\ref{fig:eq} there is a region of night side APE loss associated with this feature.

While the potential energy generation from both the radiative and frictional heating will be weighted by the local temperature through the efficiency factor $T_r/T$, the radiative heating is more strongly dependent on the temperature structure, while the frictional heating is dependent on the pattern of wind speeds.  The ``missing" potential energy generation shown in the right panel of Figure~\ref{fig:150} follows the same structure as the kinetic energy at this level.  We can see this from the flow pattern in Figure~\ref{fig:Tv}, where the strongest winds are on the night side, maximized in an equatorial jet and with significant flow in curving branches at high latitude.

\section{A Special Case: Drag in Eccentric Planet Atmospheres} \label{sec:eccentric}

Not all of the strongly irradiated close-in gas giants are on circular orbits; there are many known with significant non-zero eccentricities, the most extreme being HD 80606b, with an eccentricity of 0.934 \citep{Moutou2009}.  These eccentric planets may be subject to the same drag mechanisms that have been proposed for circular hot Jupiters, but with the added complication that the level of stellar irradiation will change throughout the orbit.  During periastron passage there will be intense stellar heating that will drive up the potential energy of the atmosphere, some of which will be converted into kinetic energy and then dissipated.  The temperature and velocity structures of the atmosphere will vary throughout the orbit.  The radiative APE generation should closely follow the temperature evolution of the atmosphere, while the frictional dissipation will follow the evolution of the kinetic energy.  At any point in the orbit there may be a mismatch between the rates of APE generation, conversion to kinetic energy, and dissipation, leading to a temporary build up of APE or kinetic energy.  This delay in energy transfer is seen in the three-dimensional model of the eccentric hot Neptune GJ436b \citep[$a=0.02872$ AU and $e=0.15$,][]{Lewis2010}, where they find that the peak in wind speeds occurs 4-8 hours after periastron passage.  The evolving dissipation may then lead to a secondary period of APE generation from the frictional heating and, depending on the delayed rates of energy transfer, this could be late enough after periastron passage that the frictional heating might dominate over the stellar heating.  As we can see, steady state assumptions and implicit time-averages are no longer valid for eccentric planet atmospheres and a more detailed analysis of the energetics will require further study.

\section{Summary and Conclusions} \label{sec:summary}

We have presented a discussion of the standard formalism for understanding atmospheric energetics, explicitly accounting for frictional heating from the dissipation of kinetic energy.  The potential energy of the atmosphere is divided into the component that can be converted into kinetic energy (the ``available" potential energy, APE) and the component associated with uniform changes in temperature (the ``unavailable" potential energy, UPE).  Uniform frictional heating will feed the UPE, while differential heating (from any source) will feed APE.  Frictional heating can be safely neglected from the energetics of the atmosphere (to some level of precision) if it is: 1) predominantly uniform, or 2) its differential heating is much less than the radiative differential heating and it contributes minimally to the generation of APE.

We give local expressions for energy generation, conversion, and dissipation in Equations (\ref{eq:finalu}-\ref{eq:finaleg}).  APE is generated in regions with positive correlations between temperature and heating (when hotter-than-average regions are heated and colder-than-average regions are cooled).  In order to determine if frictional heating matters for the energetics of an atmosphere, we must identify possible drag mechanisms and estimate their strengths and spatial variations.  We can then take global integrals of Equations (\ref{eq:finalu}-\ref{eq:finaleg}) to determine if the APE generation from frictional heating is a significant fraction of the radiative APE generation.  Given the many unknowns related to modeling hot Jupiter atmospheres, one may consider that frictional APE generation is no longer negligible once it exceeds $\sim$10\% of the radiative APE generation.  This is equivalent to saying that the heat engine efficiency of the atmosphere will be altered by more than 10\%.

As an exercise and test of this formalism, we calculated APE and UPE generation rates for numerical models that include drag as a kinetic energy sink but do not return the energy as heat.  We used the previously published numerical models by \citet{RM10} and \citet{Perna2010a}, which have identical set-ups but a range of representative drag strengths.  We found that in these models the spatial correlations between temperature and heating rates are such that in all cases the frictional heating would have worked to decrease the global generation of APE (or, equivalently, the atmosphere's heat engine efficiency), at a factor of of 1, 7, and 27\% for the models with weak, medium, and strong drag, respectively.  This provides an estimate of the drag strength at which frictional heating can significantly alter the energetics of the circulation, with the caveat that this result is strongly dependent on the spatial form we chose for the applied drag.  In these models drag timescales were horizontally constant; drag mechanisms that have greater spatial variation can have more effect on the energetics at weaker drag strengths.

Through this analysis we were also able to estimate the rate of numerical dissipation in these models.  Any atmospheric dissipation of kinetic energy (numerical or physical) must be balanced by a non-zero global heating rate, in order to generate the APE that is converted to kinetic energy (Equation~\ref{eq:balance}).  In these models the only source of heating is radiative and so this heating must be non-zero, although it means that the atmosphere is not in global radiative equilibrium.  Comparing the net radiative heating rate to a representative value for the incident stellar flux, we estimate that the rate of numerical kinetic energy loss from these models is substantial, at $\sim 5-15$\%.

By calculating local APE generation throughout these models we were able to identify the regions that contributed the most to the global energy rates.  The source term for APE in Equation (\ref{eq:finala}) is a function of the local heating rate and an efficiency factor based on the local temperature structure, $1-T_r/T$, where $T_r$ is a mass-weighted harmonic mean of temperature throughout the entire atmosphere (Equation~[\ref{eq:Tr}]).  The deeper layers of the atmosphere, which contain most of the mass, are generally hotter than the upper atmosphere.  In these models $T_r/T >1$ for pressures less than $\sim$10 bar, the same region in which all of the stellar radiative heating occurs.  This means that at these pressures APE is generated radiatively where the atmosphere is cooled (mostly on the night side) and is lost where the atmosphere is heated (mostly on the day side).

In these models most of the kinetic energy resides on the night side and---with our choice of a horizontally constant drag---this means that most of the dissipation occurs on the night side.  Since frictional dissipation always produces heating (never cooling), in this model frictional heating always decreases the APE at pressures less than 10 bar.  Under our assumption that any lost kinetic energy is locally returned as heating, this means that the APE loss due to friction is spatially correlated with the regions of strong kinetic energy and strong radiative APE generation.

Finally, we discuss how the energetics will change for planets on eccentric orbits.  Since the temperatures and wind structure of the atmosphere should vary throughout the planet's orbit, we expect evolving rates of APE production, conversion, and dissipation of kinetic energy.  A mismatch, or time delay, between these energy transfer rates would result in a build up of APE or kinetic energy, and following periastron passage there could be a secondary period of heating from frictional dissipation.

Through this analysis we have shown how frictional heating would alter the generation of available potential energy and have argued that if this effect is substantial, it could change the nature of the atmospheric circulation, since the APE is converted into kinetic energy (in the form of winds).  However, it is difficult to predict what effect we would see in the temperature and wind structure of the atmosphere.  In order to test how big of a change would result from the 27\% decrease in APE generation for the PMRc model, for example, we would need to run a model that explicitly included the localized heating from frictional dissipation.  While beyond the scope of this paper, this will be an interesting avenue to pursue in future work.

In conclusion, we suggest that performing an energetic analysis of atmospheric models for strongly-irradiated planets can be beneficial in several ways.  Aside from providing another tool with which to understand these exotic atmospheres, the diagnostics presented here can be used to evaluate whether particular heat sources (such as frictional dissipation) can significantly affect the atmospheric circulation and should be explicitly included in numerical models.  Additionally, this analysis provides an estimate of the amount of numerical kinetic energy loss in models.  While there should be some amount of physical dissipation in an atmosphere, we need to carefully consider whether the magnitude and spatial properties of the numerical dissipation introduce any erroneous effects, and further analysis is warranted.

\acknowledgements

We thank Adam Showman and Jeremy Goodman for valuable comments that helped improve the quality of this paper.
We thank the anonymous referee for useful feedback.  This research began while the authors were in residence at the Kavli Institute for Theoretical Physics, generously supported by the National Science Foundation under Grant No. PHY05-51164.  This work was performed in part under contract with the California Institute of Technology (Caltech) funded by NASA through the Sagan Fellowship Program.  KM recognizes support from NASA under grant no. PATM NNX11AD65G.

\end{document}